# Minimal Modification to Nosé-Hoover Barostat Enables Correct NPT Sampling

Somajit Dey

*Department of Physics, University of Calcutta, 92, A.P.C Road, Kolkata-700009*

Email: sdphys_rs@caluniv.ac.in

**Abstract:** The Nosé-Hoover dynamics for isothermal-isobaric (NPT) computer simulations do not generate the appropriate partition function for ergodic systems. The present paper points out that this can be corrected with a simple addition of a constant term to only one of the equations of motion. The solution proposed is much simpler than previous modifications done towards the same goal. The present modification is motivated by the work virial theorem, which has been derived for the special case of an infinitely periodic system in the first part of this paper.

## I. Introduction

The Nosé-Hoover (NH) dynamics [1] is a popular choice for isothermal-isobaric (NPT) computer simulations because of its simplicity, deterministic and time-reversible nature and its conservation of a quantity which provides a useful check on the accuracy of the coding. However, as originally noted by Hoover [1, 2], the appropriate phase space probability density for NPT ensemble does not remain stationary under this dynamics. As shown by Martyna et al. [3], the partition function associated with the NH dynamics (under the assumption of ergodicity) features an incorrect $1/V$ factor, where $V$ is the volume of the simulation box. Therefore, assuming ergodicity,

$$\langle P \rangle_{NH} = \langle P/V \rangle_{NPT} / \langle 1/V \rangle_{NPT} \neq \langle P \rangle_{NPT}, \quad (1)$$

where $\langle\ \rangle_{NH}$ denotes the long time average under NH dynamics, $\langle\ \rangle_{NPT}$ denotes isothermal-isobaric ensemble average and $P$ stands for the internal pressure. In other words, the NH trajectory does not sample the phase space correctly. The difference between $\langle P \rangle_{NH}$ and $\langle P \rangle_{NPT}$, however, is inconsequential for large system sizes. Incidentally, large system sizes show better mixing and chaos, and are better candidates for being ergodic. Also, such system sizes are almost always preferred, wherever possible, in order to reduce the finite size effects in computer simulations using periodic boundary conditions. Consequently, the NH dynamics is routinely chosen for thermo-barostatting in computer simulations.

Melchionna et al. [4] and Martyna et al. [3] have tried to remedy the incorrect sampling problem by modifying multiple equations of motion from the original NH dynamics. The two groups presented two different modifications both of which support the correct phase space probability density (distribution) for ergodic systems. The modifications, however, are very much ad hoc. That there already are two distinct correction schemes implies that there is no unique way to correct the NH dynamics. It will, therefore, be attractive if a very simple and minimal correction to the original NH dynamics comes out naturally from a rigorous analysis of the ensemble averages (time averages, if the system is ergodic). In other words, because the NH trajectory is required to generate the correct averages, the proper interrelation between the true averages may, in turn, suggest the most natural correction to the faulty NH dynamics. The present paper, in the following, takes just such an approach and derives a minimal correction to the NH equations of motion from an analysis of the virial theorem. The paper is divided into two parts. The first part (Sec. II) develops the virial theorem for an infinitely periodic system, which is the usual system of choice for computer simulations. This is important because the virial theorem has never really been worked out for such a system– the reason being, the absence of external forces [5, 6]. In Sec. II, however, we incorporate the external effects on a system under periodic boundary conditions by assuming an external, periodic potential that represents a virtual barostat coupled to the system. The only stipulation on this barostat potential is that it must generate the appropriate external virial. The second part (Sec. III) of this paper discusses the simplest correction to the NH dynamics as suggested by the virial theorem.

## II. Virial Theorem for an NPT System under PBC

Computer simulations are usually performed under periodic boundary condition (PBC). For an $N$ particle system inside an orthorhombic simulation cell (rectangular box) under PBC, the generalized coordinates are the particle position vectors $\mathbf{r}_i$ $(i = 1, 2, ..., N)$ and the three box lengths $L_\mu$ $(\mu = x, y, z)$. The Hamiltonian for this system is

$$H_{MD} = \sum_i \frac{(\mathbf{p}_i)^2}{2m_i} + U(\mathbf{r}_i, L_\mu), \quad (2)$$

where $\mathbf{p}_i$ and $m_i$ respectively denote the momentum and mass of particle $i$. $U$ denotes the internal potential that generates the internal forces on the particles and respects the PBC by adopting the minimum image convention [5]:

$$U(\mathbf{r}_i, L_\mu) = \sum_{\langle ij \rangle} \Phi(^{min}\mathbf{r}_{ij}), \quad (3)$$

where $\Phi(^{min}\mathbf{r}_{ij})$ denotes the pair potential for particles $i, j$ with

$$^{min}r_{ij}^\mu = r_{ij}^\mu - nint(r_{ij}^\mu / L_\mu) L_\mu. \quad (4)$$

Above, $\mathbf{r}_{ij} = \mathbf{r}_i - \mathbf{r}_j$ and $r^\mu$ denotes the $\mu$ component of $\mathbf{r}$. $\sum_{\langle ij \rangle}$ denotes the sum over all possible pairs of particles on or within the simulation box. The $nint()$ function gives the nearest integer to its argument. Note for later reference that, for

$$(\lambda - 1/2) L_\mu < r_{ij}^\mu < (\lambda + 1/2) L_\mu, \quad (5)$$

(where $\lambda$ is an integer)

$$nint(r_{ij}^\mu / L_\mu) = \lambda. \quad (6)$$

Discontinuity in the $nint()$ function appears only as $r_{ij}^\mu / L_\mu$ approaches $\lambda \pm 1/2$. By this point, however, any interaction between particles $i$ and $j$ should have decayed to zero under the minimum image convention [5]. This, of course, excludes the long range electrostatic forces which require special handling under PBC not within the scope of this paper. For short range forces, therefore,

$$F_{ij}^\mu \frac{\partial (nint(r_{ij}^\mu / L_\mu))}{\partial r_{ij}^\mu} = 0 = F_{ij}^\mu \frac{\partial (nint(r_{ij}^\mu / L_\mu))}{\partial L_\mu}, \quad (7)$$

where $\mathbf{F}_{ij}$ is the force applied on particle $i$ by particle $j$.

The suffix MD in Eq. (2) refers to the fact that this Hamiltonian generates the standard molecular dynamics (MD) that simulates isolated (constant energy, constant box lengths) systems not coupled to a barostat or thermostat. For isobaric simulations, however, the system is no more isolated because the internal energy $H_{MD}$ does not remain constant. Each box length also becomes a degree of freedom in this case. To account for these, the system Hamiltonian must be augmented. To be specific, the new Hamiltonian should be

$$H = H_{MD} + \bar{H}(\Lambda_\mu, L_\mu, \mathbf{r}_i). \quad (8)$$

where $\Lambda_\mu$ denotes the momentum conjugate to $L_\mu$. Instead of $H_{MD}$, $H$ would be conserved for isobaric dynamics. $\bar{H}$ may be considered an external potential serving as proxy for a virtual barostat. To mimic systems of volume $V$ bounded by walls applying constant pressure $\mathbf{P}$, two conditions must be satisfied:

$$\partial \bar{H} / \partial L_\mu > 0, \quad \text{for } \mathbf{P} > 0, \quad (9)$$

$$G_{\text{ext}} = \sum_i \mathbf{r}_i \cdot \vec{\nabla}_{\mathbf{r}_i} \bar{H} + \sum_\mu L_\mu \frac{\partial \bar{H}}{\partial L_\mu} = 3\mathbf{P}V. \quad (10)$$

$G_{\text{ext}}$ denotes the generalized external virial. $\bar{H}$ is also required to comply with the PBC.

Let us now couple a thermostat to our already barostatted system described by $H$. For this thermostatted system, the generalized equipartition theorem [7] states

$$\left\langle \theta_n \frac{\partial H}{\partial \theta_l} \right\rangle = \delta_{nl} k \mathbf{T}, \quad (11)$$

where $\theta_l$ refers to any generalized coordinate or momentum, $k$ denotes the Boltzmann constant, $\mathbf{T}$ is the thermostat temperature and $\langle\ \rangle$ denotes the canonical ensemble average. The proof of Eq. (11) is trivial and can be looked up in any textbook derivation of the equipartition and virial theorems [8]. The proof, however, requires $\exp(-H/k\mathbf{T})$ to be zero whenever any generalized coordinate or momentum assumes an extreme value. This, therefore, needs to be established for our infinitely periodic system, if Eq. (11) is to hold. For any momentum taking extreme value, $\exp(-H/k\mathbf{T})$ is trivially zero. For a particle sitting on any of the box edges, its periodic image sits on the opposite edge of the simulation box. According to Eq. (4) then the particle and its image should interact as effectively coincident particles, which makes $U$ infinite and thus, $\exp(-H/k\mathbf{T}) = 0$. Lastly, by virtue of Eq. (9), $L_\mu \to \infty$ implies $\bar{H} \to \infty$, which in turn implies $\exp(-H/k\mathbf{T}) \to 0$.

Using Eq. (11), we get the standard equipartition theorem,

$$\left\langle \sum_i \frac{(\mathbf{p}_i)^2}{m_i} \right\rangle = Xk\mathbf{T}, \quad (12)$$

and the generalized virial theorem,

$$\left\langle \sum_i \mathbf{r}_i \cdot \vec{\nabla}_{\mathbf{r}_i} H + \sum_\mu L_\mu \frac{\partial H}{\partial L_\mu} \right\rangle = (X+3)k\mathbf{T}, \quad (13)$$

$X$ being the number of independent particle coordinates. Using Eq. (12), we rewrite Eq. (13) as

$$\langle G_{\text{ext}} + G_{\text{int}} \rangle = \left\langle \sum_i \frac{(\mathbf{p}_i)^2}{m_i} \right\rangle + 3k\mathbf{T}, \quad (14)$$

where $G_{\text{int}}$ denotes the generalized internal virial,

$$G_{\text{int}} = \sum_i \mathbf{r}_i \cdot \vec{\nabla}_{\mathbf{r}_i} U + \sum_\mu L_\mu \frac{\partial U}{\partial L_\mu}. \quad (15)$$

Using Eq. (3), (4) and (7), the right hand side of Eq. (15) turns out to be $-\sum_{\langle ij \rangle}{}^{\min} \mathbf{r}_{ij} \cdot \mathbf{F}_{ij}$. Using Eq. (10) and rearranging, Eq. (14) therefore becomes

$$\langle 3\mathbf{P}V \rangle = \left\langle \sum_i \frac{(\mathbf{p}_i)^2}{m_i} + \sum_{\langle ij \rangle}{}^{\min} \mathbf{r}_{ij} \cdot \mathbf{F}_{ij} \right\rangle + 3k\mathbf{T}. \quad (16)$$

Now, the internal pressure, as a function of the microstate, is conventionally defined as [9]

$$P = \frac{1}{3V}\left( \sum_i \frac{(\mathbf{p}_i)^2}{m_i} + \sum_{\langle ij \rangle}{}^{\min} \mathbf{r}_{ij} \cdot \mathbf{F}_{ij} \right). \quad (17)$$

This definition works because under the isothermal-isobaric partition function, $P$ averages to the external pressure $\mathbf{P}$ (Appendix A). Eq. (16), therefore, implies

$$\langle PV \rangle_{\text{NPT}} - \mathbf{P} \langle V \rangle_{\text{NPT}} + k\mathbf{T} = 0. \quad (18)$$

This is known as the work virial theorem. Under the assumption of ergodicity, therefore, any dynamics that samples the correct phase space distribution for our isothermal-isobaric system under PBC should reproduce the above equality between the corresponding time averages. It is clear from the above development that the $k\mathbf{T}$ term in the work virial theorem comes due to the box length degrees of freedom. Although we have focused exclusively on infinitely periodic systems above, Eq. (18) is valid for other systems and can be derived from general statistical mechanics using the thermodynamic definition of internal pressure, as shown in Appendix B of Ref. [3]. Also note that in all of the above, no assumption has been made as to the actual form of the external potential $\bar{H}$, apart from its periodicity and that it satisfies Eq. (9) and (10). By way of illustration, however, Appendix B discusses the barostatting mechanics for a simple choice of $\bar{H}$.

### III. Correction to Nosé-Hoover (NH) Dynamics

To see what correction to the NH dynamics is suggested by Eq. (18), let us now consider the original NH equations of motion. Written in terms of $\mathbf{r}_i$, $\mathbf{p}_i$, $L_\mu$ the equations are:

$$\dot{\mathbf{r}}_i = \frac{\mathbf{p}_i}{m_i} + \nu_P \eta \mathbf{r}_i, \quad (19)$$

$$\dot{\mathbf{p}}_i = \mathbf{F}_i - (\nu_T \zeta + \nu_P \eta) \mathbf{p}_i, \quad (20)$$

$$\frac{\dot{L}_\mu}{L_\mu} = \nu_P \eta, \quad (21)$$

$$\dot{\zeta} = \frac{\nu_T}{Q_T}\left( \sum_i \frac{(\mathbf{p}_i)^2}{m_i} - Xk\mathbf{T} \right), \quad (22)$$

$$\dot{\eta} = \frac{3\nu_P}{Q_P}(P - \mathbf{P})V. \quad (23)$$

Above, $\mathbf{F}_i$ is the total internal force on particle $i$. The parameters $\nu_P$ and $\nu_T$ are of dimension $[Time]^{-1}$ and govern the rate of coupling of the particles to the barostat and the thermostat respectively. Making them zero uncouples the barostat and thermostat which, in turn, takes away all the external forces operating on the system. The equations of motion for this isolated system then reduce to Newtonian dynamics. $\zeta$ is called the friction coefficient and $\eta$ is called the strain rate. It may be noted that Eq. (21) actually encodes 3 equations of motion whereas the original NH dynamics only had one equation for the rate of volume fluctuation. This is because the original NH dynamics [1] assumed a cubic simulation box where knowledge of the box volume was sufficient to determine the box dimension. Our formulation of the NH dynamics, however, is not confined to the special case of a cubic simulation box. Since the same strain rate is used for all the directions, volume fluctuation is isotropic in NH dynamics. This implies $\frac{\dot{L}_x}{L_x} = \frac{\dot{L}_y}{L_y} = \frac{\dot{L}_z}{L_z}$.

Using this along with the original NH equation of motion $\dot{V} = 3\eta V$, one obtains Eq. (21). The primary role of $Q_T$ and $Q_P$ in Eq. (22) and (23) is to make $\dot{\zeta}$ and $\dot{\eta}$ dimensionless. They may also be chosen to scale with the system size, one choice being $Q_T = Xk\mathbf{T}$ and $Q_P = (X+3)k\mathbf{T}$. The NH dynamics with equations of motion (19)-(23) conserves the quantity

$$C = E + \mathbf{P}V + \frac{1}{2}(Q_T \zeta^2 + Q_P \eta^2) + Xk\mathbf{T}\xi, \quad (24)$$

where $E = \sum_i (\mathbf{p}_i \cdot \mathbf{p}_i / 2m_i) + U$ and $\xi(t) = v_T \int_0^t \zeta(t') dt'$.

As mentioned in Sec. I, the NH barostat does not reproduce the correct NPT average for the internal pressure. So let us now focus on the only equation of motion in NH dynamics that features both $P$ and $\mathbf{P}$, viz. Eq. (23). Time averaging both sides of that equation gives

$$\langle PV \rangle_t - \mathbf{P}\langle V \rangle_t = \frac{Q_P}{3v_P} \langle \dot{\eta} \rangle_t = \frac{Q_P}{3v_P}\left[\frac{\eta(t)-\eta(0)}{t}\right]. \quad (25)$$

When $t$ is large, the last term in Eq. (25) tends to zero because $\eta$ remains bound in order to keep $C$ conserved. Therefore,

$$\langle PV \rangle_{NH} - \mathbf{P}\langle V \rangle_{NH} = 0. \quad (26)$$

Note that this equation differs from the correct work virial theorem for the NPT ensemble, viz. Eq. (18), only by a $k\mathbf{T}$ term. The simplest, most natural correction to the NH dynamics would therefore amount to changing the equation of motion of $\eta$ from Eq. (23) to

$$\dot{\eta} = \frac{3v_P}{Q_P}\left[(P-\mathbf{P})V + k\mathbf{T}\right]. \quad (27)$$

The corrected NH dynamics, viz. Eq. (19)-(22) and (27), conserves the quantity

$$\tilde{C} = C - k\mathbf{T}\ln V. \quad (28)$$

This correction to the strain rate equation of motion, viz. replacing Eq. (23) with Eq. (27), is the only modification needed in the NH dynamics to generate the correct partition function under ergodicity, as will be shown below. This makes sense, because the corrected dynamics takes into account both of the two key theorems regarding the NPT ensemble – Eq. (22) ensures the equipartition theorem and Eq. (27), the work virial theorem.

Let us now derive the partition function generated by the corrected dynamics, when ergodic. The Jacobian $J(t)$ associated with the volume element of the extended system phase space satisfies [3]

$$dJ/dt = -J\left[\sum_\Omega \partial\dot{\Omega}/\partial\Omega\right], \quad (29)$$

where $\Omega$ denotes the independent elements of the set $(\mathbf{r}_i, \mathbf{p}_i, L_\mu, \eta, \xi, \zeta)$. Solving Eq. (29),

$$J(t) = \exp(X\xi)/(L_x L_y L_z) = \exp(X\xi - \ln V). \quad (30)$$

The partition function associated with an ergodic dynamics that samples the phase space under the sole constraint, $\tilde{C} = \Xi$, $\Xi$ being any constant, is given by

$$\Delta = \int d\zeta d\eta d\xi dL_x dL_y dL_z \int_{D(L_\mu)} d\omega d\pi \left[J\delta(\tilde{C}-\Xi)\right], \quad (31)$$

where $d\omega = \prod_{j=1}^{X/3}(dr_j^x dr_j^y dr_j^z)$, $d\pi = \prod_{j=1}^{X/3}(dp_j^x dp_j^y dp_j^z)$ and $D(L_\mu)$ denotes the domain defined by the box lengths. Using $\tilde{C}$ from Eq. (28) and a standard $\delta$ function identity, we get

$$\delta(\tilde{C}-\Xi) = \frac{1}{Xk\mathbf{T}}\delta(\xi-\xi_0), \quad (32)$$

where

$$X\xi_0 = \ln V + \frac{\Xi - \left\{E + \mathbf{P}V + \frac{1}{2}(Q_T\zeta^2 + Q_P\eta^2)\right\}}{k\mathbf{T}}.$$

Putting $J$ given by Eq. (30), along with the right hand side of Eq. (32), in the integrand of Eq. (31) and integrating over $\xi$,

$$\Delta = \frac{1}{Xk\mathbf{T}}\int d\zeta d\eta dL_x dL_y dL_z \int_{D(L_\mu)} d\omega d\pi \left[e^{X\xi_0 - \ln V}\right]. \quad (33)$$

Integrating now over $\zeta$ and $\eta$,

$$\Delta = \frac{2\pi e^{\frac{\Xi}{k\mathbf{T}}}}{Xk\mathbf{T}\sqrt{Q_T Q_P}}\int dL_x dL_y dL_z \int_{D(L_\mu)} d\omega d\pi \left[e^{-(E+\mathbf{P}V)/k\mathbf{T}}\right], \quad (34)$$

which is the correct NPT partition function within a constant factor.

For completeness, let us also show that the NPT phase space probability density,

$$f_{NPT} \propto e^{-\left[E+\mathbf{P}V+\frac{1}{2}(Q_T\zeta^2+Q_P\eta^2)\right]/k\mathbf{T}}, \quad (35)$$

is a stationary solution to the probability flow equation,

$$\frac{\partial f}{\partial t} + \sum_\Omega \dot{\Omega}\frac{\partial f}{\partial \Omega} + f\sum_\Omega \frac{\partial\dot{\Omega}}{\partial \Omega} = 0. \quad (36)$$

In what follows, we write down all the terms in Eq. (36) as applicable for $f_{NPT}$ given by Eq. (35).

$$\sum_i \dot{\mathbf{r}}_i \cdot \vec{\nabla}_{\mathbf{r}_i} f + \sum_\mu \dot{L}_\mu \frac{\partial f}{\partial L_\mu} = \left[\sum_i \frac{\mathbf{p}_i \cdot \mathbf{F}_i}{m_i} + v_P\eta\left(\sum_{\langle ij \rangle}^{\min} \mathbf{r}_{ij}\cdot\mathbf{F}_{ij} - 3PV\right)\right]\frac{f}{k\mathbf{T}}$$

$$\sum_i \dot{\mathbf{p}}_i \cdot \vec{\nabla}_{\mathbf{p}_i} f = -\left(\sum_i \frac{\mathbf{p}_i\cdot\mathbf{F}_i}{m_i} - (v_T\zeta + v_P\eta)\sum_i \frac{(\mathbf{p}_i)^2}{m_i}\right)\frac{f}{k\mathbf{T}}$$

$$\dot{\zeta}\frac{\partial f}{\partial \zeta} = -v_T\zeta\left(\sum_i \frac{(\mathbf{p}_i)^2}{m_i} - Xk\mathbf{T}\right)\frac{f}{k\mathbf{T}}$$

$$\dot{\eta}\frac{\partial f}{\partial \eta} = -v_P\eta\left(\sum_i \frac{(\mathbf{p}_i)^2}{m_i} + \sum_{\langle ij \rangle}^{\min}\mathbf{r}_{ij}\cdot\mathbf{F}_{ij} - 3\mathbf{P}V + 3k\mathbf{T}\right)\frac{f}{k\mathbf{T}}$$

$$f\left(\sum_i(\vec{\nabla}_{\mathbf{r}_i}\cdot\dot{\mathbf{r}}_i + \vec{\nabla}_{\mathbf{p}_i}\cdot\dot{\mathbf{p}}_i) + \sum_\mu \frac{\partial \dot{L}_\mu}{\partial L_\mu}\right) = (-Xv_T\zeta + 3v_P\eta)f$$

(37)

All terms on the right hand sides of equations (37), taken together, cancel each other. Therefore, $f_{NPT}$ is indeed the equilibrium solution of the continuity equation [Eq. (36)].

### IV. Conclusion

For ergodic systems, therefore, our corrected NH dynamics gives trajectory (time) averages that accurately represent the corresponding isothermal-isobaric ensemble average for state functions, viz. variables that depend only on the box dimensions and the particle positions and momenta. This implies that the desired equality between the average internal pressure and the external (barostat) pressure is automatically satisfied. Note that Eq. (27) can be derived by demanding $f_{NPT}$, as given by Eq. (35), be a stationary solution of Eq. (36) under the equations of motion: Eq. (19)-(22). This implies that our correction to $\dot{\eta}$ is unique. In other words, no other correction to the strain rate equation of motion can generate the correct NPT distribution if all other NH equations of motion are kept unmodified.

Note that the original strain rate equation [Eq. (23)] differs from its proposed replacement [Eq. (27)] only by a constant term, viz. $3v_P k\mathbf{T}/Q_P$. With such a minimal modification to only one equation of motion (that of the strain rate), the NH dynamics can thus sample the correct NPT ensemble. This correction, therefore, maintains the essential simplicity of the original NH equations of motion. In contrast, the modifications due to Melchionna et al. [4] and Martyna et al. [3] changed multiple equations of motion from the original NH set in order to generate the correct NPT sampling.

Computer simulations are usually initialized with the centre of mass (c.m.) of the system stationary at the origin, which is usually at the centre of the simulation box. When $\sum_i \mathbf{F}_i = 0$, the NH dynamics, including our proposed correction, conserve the total momentum under such initializations, thus retaining the c.m. at the origin for the entire time evolution of the system. The states of only $N-1$ of the $N$ particles are independent in this case. The formalism presented in this paper remains perfectly valid for this case with $X = 3N - 3$. It may be remarked for comparison that the modified dynamics proposed by Martyna et al. [3] also shows this behaviour as discussed in detail in Ref. [10].

In closing, we note that the original NH dynamics [1] considered a cubic simulation box. Starting with a cubic box, i.e. $L_x = L_y = L_z$, Eq. (21) implies that all box lengths remain equal along the trajectory − which is to be expected because volume is fluctuating isotropically. For this case, Eq. (21) can be replaced by

$$\dot{L} = v_P \eta L, \tag{38}$$

where $L$ is the length of the cubic simulation box.

## Acknowledgement

The author is funded by the Council of Scientific and Industrial Research, India through a research fellowship [File No. 09/028(0960)/2015-EMR-I].

## Appendix A

From Eq. (15) and (17), we can write the internal pressure as

$$P = \frac{1}{3V}\left(\sum_i \frac{(\mathbf{p}_i)^2}{m_i} - \sum_i \mathbf{r}_i \cdot \vec{\nabla}_{\mathbf{r}_i} U - \sum_\mu L_\mu \frac{\partial U}{\partial L_\mu}\right). \tag{39}$$

Averaging under the NPT ensemble,

$$\langle P \rangle = \left\langle \frac{1}{3V}\left(\sum_i \frac{(\mathbf{p}_i)^2}{m_i} - \sum_i \mathbf{r}_i \cdot \vec{\nabla}_{\mathbf{r}_i} U\right)\right\rangle - \sum_\mu \left\langle \frac{L_\mu}{3V} \frac{\partial U}{\partial L_\mu}\right\rangle. \tag{40}$$

As in Eq. (11), for an ensemble with probability density $\propto \exp(-E/k\mathbf{T})$,

$$\left.\begin{array}{c}\int (\mathbf{p}_i \cdot \vec{\nabla}_{\mathbf{p}_i} E) e^{-E/k\mathbf{T}} d\omega d\pi \\ \int (\mathbf{r}_i \cdot \vec{\nabla}_{\mathbf{r}_i} E) e^{-E/k\mathbf{T}} d\omega d\pi\end{array}\right\} = 3k\mathbf{T}\int e^{-E/k\mathbf{T}} d\omega d\pi, \tag{41}$$

where $d\omega = \prod_{j=1}^{X/3}(dr_j^x dr_j^y dr_j^z)$ and $d\pi = \prod_{j=1}^{X/3}(dp_j^x dp_j^y dp_j^z)$. Using the equalities in Eq. (41), the first term on the right hand side of Eq. (40) can be easily shown to be zero. Now,

$$\left\langle \frac{L_x}{V}\frac{\partial U}{\partial L_x}\right\rangle = \frac{\int_0^\infty dL_z dL_y dL_x \frac{e^{-\mathbf{P}V/k\mathbf{T}}}{L_y L_z} \int_{D(L_\mu)} \frac{\partial U}{\partial L_x} e^{-U/k\mathbf{T}} d\omega}{\int_0^\infty dL_z dL_y dL_x \int_{D(L_\mu)} e^{-(\mathbf{P}V+U)/k\mathbf{T}} d\omega} \tag{42}$$

where $D(L_\mu)$ denotes the domain defined by the box lengths. Using differentiation under the integral sign,

$$\int_{D(L_\mu)} \frac{\partial U}{\partial L_x} e^{-U/k\mathbf{T}} d\omega = -k\mathbf{T}\frac{\partial}{\partial L_x}\int_{D(L_\mu)} e^{-U/k\mathbf{T}} d\omega. \tag{43}$$

By integration by parts,

$$\int_0^\infty dL_x e^{\frac{-\mathbf{P}V}{k\mathbf{T}}} \frac{\partial}{\partial L_x}\int_{D(L_\mu)} e^{\frac{-U}{k\mathbf{T}}} d\omega = \frac{\mathbf{P}L_y L_z}{k\mathbf{T}}\int_0^\infty dL_x \int_{D(L_\mu)} e^{\frac{-(\mathbf{P}V+U)}{k\mathbf{T}}} d\omega. \tag{44}$$

Using Eq. (43) and (44) in the numerator of Eq. (42), we ultimately obtain

$$\left\langle \frac{L_x}{V}\frac{\partial U}{\partial L_x}\right\rangle = -\mathbf{P}. \tag{45}$$

Similarly for $y$ and $z$. From Eq. (40), therefore,

$$\langle P \rangle = -\sum_\mu \left\langle \frac{L_\mu}{3V}\frac{\partial U}{\partial L_\mu}\right\rangle = \mathbf{P}. \tag{46}$$

## Appendix B

For

$$\bar{H}(\Lambda_\mu, L_\mu) = \sum_\mu \frac{(\Lambda_\mu)^2}{2Q_\mu} + \mathbf{P}V, \tag{47}$$

(where $Q_\mu$ is a mass parameter) the equations of motion generated by the Hamiltonian $H$ [Eq. (8)] are

$$\dot{\mathbf{r}}_i = \mathbf{p}_i/m_i; \qquad \dot{\mathbf{p}}_i = -\vec{\nabla}_{\mathbf{r}_i} U = \mathbf{F}_i; \tag{48}$$

$$\dot{L}_\mu = \Lambda_\mu/Q_\mu; \qquad \dot{\Lambda}_\mu = -\mathbf{P}V/L_\mu - \partial U/\partial L_\mu$$
$$= -\frac{\mathbf{P}V}{L_\mu} - \sum_{\langle ij \rangle} nint\left(\frac{r_{ij}^\mu}{L_\mu}\right) F_{ij}^\mu. \tag{49}$$

To see exactly how these equations give the isobaric dynamics, let us probe the last equation of motion for $\mu = x$. It can be rewritten as

$$\dot{\Lambda}_x = (P_{xx} - \mathbf{P})L_y L_z - \frac{1}{L_x}\sum_i \left(r_i^x F_i^x + (p_i^x)^2/m_i\right). \tag{50}$$

$P_{\mu\mu}$ denotes a diagonal component of the internal pressure tensor,

$$P_{\mu\mu} = \frac{1}{V}\left[\sum_{\langle ij \rangle}^{min} r_{ij}^\mu F_{ij}^\mu + \sum_i (p_i^\mu)^2/m_i\right].$$

Trajectory or time average of Eq. (50) gives

$$\langle \dot{\Lambda}_x \rangle_t = \langle (P_{xx} - \mathbf{P}) A_{yz} \rangle_t - \left\langle \frac{1}{L_x}\sum_i \left(r_i^x F_i^x + \frac{(p_i^x)^2}{m_i}\right)\right\rangle_t. \tag{51}$$

$A_{yz} = L_y L_z$ is the area of each of the two faces of the simulation box that are parallel to the $y-z$ plane. Below, these box surfaces will be referred to as $yz$ faces. $\Lambda_x$ remains bounded along the trajectory because the dynamics conserves $H$. Therefore,

$$\langle \dot{\Lambda}_x \rangle_t = \lim_{t\to\infty}\frac{\Lambda_x(t) - \Lambda_x(0)}{t} = 0.$$

Under the assumption of ergodicity, the last term in Eq. (51) can be written as

$$-\sum_E \sum_{L_\mu}\left[\left\langle \sum_i \left(r_i^x F_i^x + \frac{(p_i^x)^2}{m_i}\right)\right\rangle_{H_{MD}=E}^{L_\mu} \frac{\tau(L_\mu, E)}{L_x}\right] = 0, \tag{52}$$

where $\langle\ \rangle_{H_{MD}=E}^{L_\mu}$ denotes the trajectory average obtained from standard molecular dynamics with energy, $E$ and box lengths, $L_\mu$. $\tau(L_\mu, E)$ denotes the fraction of the total time that the system spends in the macrostate defined by $H_{MD} = E$ and box dimensions $= L_\mu$. The equality in Eq. (52) then follows from the virial theorem of classical mechanics [11], viz.

$$\left\langle \sum_i \frac{(p_i^x)^2}{m_i} + \sum_i r_i^x F_i^x\right\rangle_{MD} = \left\langle \frac{d}{dt}\sum_i r_i^x p_i^x\right\rangle_{MD},$$

$$= \lim_{t\to\infty}\frac{\sum_i r_i^x(t) p_i^x(t) - \sum_i r_i^x(0) p_i^x(0)}{t} = 0$$

by virtue of the boundedness of $r_i^x$ and $p_i^x$. From Eq. (51), therefore,

$$\langle P_{xx} A_{yz} \rangle_t = \mathbf{P} \langle A_{yz} \rangle_t. \qquad (53)$$

Eq. (53) gives the essential force balance that equates the average internal force on a $yz$ face to the average external force on that face. Defining average internal normal pressure as the average internal force on a surface divided by the average surface area, we get from Eq. (53)

$$P_x^N = \mathbf{P},$$

where $P_x^N$ represents the average internal normal pressure along the $x$ axis. Likewise for $\mu = y, z$. Thus, barostatting is achieved.

To achieve isothermal-isobaric (NPT) dynamics, a Nosé-Hoover thermostat [1, 12] may be coupled to the isobaric dynamics discussed above. However, different temperature control variables should preferably be used for the particle and box length degrees of freedom [13]. Such a thermostatted isobaric dynamics gives the correct NPT behaviour. This has been verified by the present author both analytically and numerically, and will be reported elsewhere.

### References

1. W. G. Hoover, *Phys. Rev. A* **31**, 1695 (1985).
2. W. G. Hoover, *Phys. Rev. A* **34**, 2499 (1986).
3. G. J. Martyna, D. J. Tobias, M. L. Klein, *J. Chem. Phys.* **101**, 4177 (1994).
4. S. Melchionna, G. Ciccotti, B. L. Holian, *Mol. Phys.* **78**, 533 (1993).
5. M. P. Allen, D. J. Tildesley, *Computer Simulation of Liquids* (Clarendon Press, Oxford, 1987).
6. J. J. Erpenbeck, W. W. Wood, in *Statistical Mechanics B. Modern Theoretical Chemistry*, edited by B. J. Berne (Plenum, New York, 1977).
7. A. Münster, *Statistical Thermodynamics* (Springer, Berlin, 1969).
8. R. K. Pathria, P. D. Beale, *Statistical Mechanics* (Academic Press, Oxford, 2011).
9. H. C. Anderson, *J. Chem. Phys.* **72**, 2384 (1980).
10. M. E. Tuckerman, Y. Liu, G. Ciccotti, G. J. Martyna, *J. Chem. Phys.* **115**, 1678 (2001).
11. H. Goldstein, C. P. Poole, J. L. Safko, *Classical Mechanics* (Pearson, 2002).
12. See Appendix of [B. L. Holian, A. F. Voter, R. Ravelo, *Phys. Rev. E* **52**, 2338 (1995)] for a leap-frog integrator.
13. S. Nosé, *Mol. Phys.* **57**, 187 (1986).